\documentclass[10pt,letter]{article}
\usepackage[latin1]{inputenc}
\usepackage{amsmath,amsfonts,amssymb,graphicx}
\usepackage{authblk}

\begin{document}

\title{\textbf{A statistical method for analyzing and comparing spatiotemporal cortical activation patterns}}
\author[1,2]{Patrick Krauss}
\author[2]{Claus Metzner}
\author[1,2]{Achim Schilling}
\author[1]{\\ Konstantin Tziridis}
\author[3]{Maximilian Traxdorf}
\author[1]{Holger Schulze*}
\affil[1]{\small{Experimental Otolaryngology, University Hospital Erlangen, Friedrich-Alexander University Erlangen-Nürnberg (FAU), Germany}}
\affil[2]{Department of Physics, Biophysics Group, Friedrich-Alexander University Erlangen-Nürnberg (FAU), Germany}
\affil[3]{Department of Otorhinolaryngology, Head and Neck Surgery, University Hospital Erlangen, Germany}
\affil[ ]{ }
\affil[ ]{\large{\textit{* holger.schulze@uk-erlangen.de}}}
\renewcommand\Authands{ and }

%\twocolumn[\begin{@twocolumnfalse} 

\maketitle

%\section*{Abstract}

\begin{abstract}\textbf{We present a new statistical method to analyze multichannel steady-state local field potentials (LFP) recorded within different sensory cortices of different rodent species. Our spatiotemporal multi-dimensional cluster statistics (MCS) method enables statistical analyzing and comparing clusters of data points in n-dimensional space. We demonstrate that using this approach stimulus-specific attractor-like spatiotemporal activity patterns can be detected and be significantly different from each other during stimulation with long-lasting stimuli. Our method may be applied to other types of multichannel neuronal data, like EEG, MEG or spiking responses and used for the development of new read-out algorithms of brain activity and by that opens new perspectives for the development of brain-computer interfaces.} 
\end{abstract}

%\end{@twocolumnfalse}]

\section*{Introduction}

Brains use sensory systems to generate internal representations of the world. These internal representations serve animal organisms as a frame of reference to guide their behavior. Nevertheless, these internal representations of the sensory surround are not a mere copy of the world, but rather models that represent certain aspects of it that are selected by sensory and cognitive filters. The available information is thereby reduced to a fraction momentarily considered as relevant, and only this part may then be perceived.

Central sensory systems realize this complex task by a combination of parallel and sequential processing of sensory information within neuronal networks, in interaction with cognitive, affective and motivational systems. It seems common sense that the internal sensory representations that result from such processing are some kind of spatiotemporal cortical activation patterns. In particular, spatiotemporal patterns of neuronal activity distributed over large numbers of neurons have been proposed (population coding) that are able to form attractor-like non-linear dynamics (Dehaene et al., 2003; Daelli and Treves, 2010). In order to form such attractor dynamics, correlated activity in only a subset of neurons within a sensory field is sufficient, whereby the mean discharge rate does not even have to be above that of spontaneous activity: just the spatiotemporal pattern of activity would be different compared to spontaneous activity without a perceptual relevance (Kumar et al., 2008; Ringach, 2009; Tomov et al., 2014). Although a number of studies have described such attractor-like dynamics, e.g. in the auditory cortex, based on electrocorticogram (e.g., Ohl et al., 2001; Ohl et al., 2003a, b; Deliano et al., 2009) or extracellular spike recordings (Harris et al., 2011), most of these studies usually analyze stimulus evoked responses within the first hundred milliseconds after stimulus onset in comparison to spontaneous activity (e.g. Harris et al., 2011) or during certain marked states that correlate with certain behavioral states (cf. Ohl et al., 2001). As a stimulus like a tone for example may be perceived as long as it is present, the internal representation of that stimulus must be encoded somehow in the ongoing steady-state activity within sensory cortex. There are only very few studies investigating responses to long-term stimuli at all (e.g., Goldberg et al., 1964), and it seems that at least some neurons might be able to respond tonically to even such long-lasting stimuli (Javel, 1996). Nevertheless, as large-scale multichannel recordings were not possible by that time, these classical studies describe single unit spiking activity only and therefore could not analyze spatiotemporal activity patterns. The reason why steady-state attractor dynamics still have not been analyzed yet may be due to the fact that proper statistics that allow for a differentiation between steady-state spatiotemporal patterns belonging to the same or different internal representations have not been developed yet.

In this report, we present a new statistical method to analyze multichannel steady-state local field potentials (LFP) recorded within auditory cortex of Mongolian gerbils. We demonstrate that using this approach stimulus-specific spatiotemporal activity patterns can be detected and be significantly distinguished from each other during stimulation with long-lasting stimuli.

\section*{Methods}

\paragraph*{Data acquisition}
Mongolian gerbils (\textit{Meriones unguiculatus}) were housed in standard animal racks (Bio A.S. Vent Light, Zoonlab GmbH, Castrop-Rauxel, Germany) in groups of 2 animals per cage with free access to water and food at $20-24^{\circ}C$  room temperature under 12/12 h dark/light cycle.
 
Gerbils were purchased from Charles River Laboratories Inc. (Sulzfeld, Germany).

Electrophysiological multichannel recordings were made in rodent sensory cortex, namely in auditory cortex of 15 adult male Mongolian gerbils.

Gerbils were deeply anesthetized by a mixture of ketamine, xylacine, isotonic NaCl solution, and atropine at a mixing ratio of 9:1:8:2. Initial dose was 0.3 ml s.c., and anesthesia was maintained via s.c. infusion of the anesthetic solution, continuously applied by a syringe pump at a rate of 0.2 to 0.3 ml/h. During surgery and recording, the animal's body temperature was kept constant at $37^{\circ}C$  by a remote controlled heating pad (FHC Inc., Bowdoin, ME, USA).

For recordings in gerbil auditory cortex, the skin over the skull and the musculature covering the temporal bone on the left (recording) side were partly removed. The auditory cortex was then exposed by craniotomy, leaving the dura intact. Finally, a stainless steel screw was fixed to the frontal bones with dental acrylic and served as a head anchor for stereotaxic fixation. After surgery, the still anaesthetized animals were transferred into an electrically and acoustically shielded, anechoic recording chamber.

Acute electrophysiological multi-channel recordings were made using 16-channel arrays (geometry $4 \times 4$, spacing $250 \mu m$, singe electrodes with $2 M \Omega$ impedance, $2 \mu m$ tip diameter, Clunbury Scientific, USA) inserted into the auditory cortex. LFPs were recorded (1 kHz sampling rate, filtering: 50 Hz notch and low pass with 200 Hz cut-off frequency, amplification factor: 20000) during free field sound stimulation with pure tones (70 dB SPL) of different frequencies (1, 2, and 4 kHz) To obtain sustained LFP activity, recordings during stimulation were carried out continuously for 3 minutes. In addition, spontaneous activity without any stimulation was recorded for the same amount of time.

\paragraph*{Data visualization}
Data are visualized using multidimensional scaling (MDS) which can be used to project data points from a n-dimensional space onto a lower m-dimensional target space, such that all mutual Euclidean distances in n-dimensional space are preserved in target space. By that, distance in n- or m-dimensional space is a measure of dissimilarity between data points (cf. Fig. 1 and 2) which in the context of this report refers to spatiotemporal patterns of neuronal activity (cf. Fig. 3). Note that MDS in this report is not used as an analytical tool, but for visualization of high-dimensional data in two-dimensional data plots only. All statistics developed and presented below operate in n-dimensional space and are completely independent of MDS.

\paragraph*{Ethics statements}
The use and care of animals was approved by the state of Bavaria (Regierungspräsidium Mittelfranken, Ansbach, Germany; AZ: 54-2532.1-02/13).

\section*{Results}
In this Results section we will first theoretically describe our approach to statistically compare clusters of data points in n-dimensional space (Fig. 1 and 2). We then apply the method to LFP data from gerbil auditory cortex to demonstrate that the method is able to distinguish between different cortical LFP activation patterns (Fig. 3).

\paragraph*{Comparing two clusters in n-dimensional space}
Our spatiotemporal multidimensional cluster statistics (MCS) method enables statistical analysis and comparing clusters of data points in n-dimensional space. The different steps of the procedure are detailed in Figure 1: We start from a data matrix (Fig 1a) containing the coordinates (1 to n) and cluster labels (A, B) of all data points in n-dimensional space. For visualization, these data points are projected from n-dimensional space to two-dimensional space by MDS (Fig. 1e, cf. Methods). Next, all pairwise Euclidean distances between points are calculated (Fig. 1b, f). Then, the mean distances between intra-cluster points (d(A,A); d(B,B)) and inter-cluster points (d(A,B) = d(B,A)) are derived by averaging the Euclidean distances according to the points' cluster labels (Fig. 1c). From these the mean distances (referred to as proximities, Fig. 1g, large circular areas refer to centroid position and diameter of clusters) a discrimination value $\Delta$ is computed (Fig. 1d) which quantifies the discriminability and density of the clusters of interest (cluster A and B in the example shown): The more negative this value is, the more distinct and/or dense the considered clusters are. Note that this analysis does not take into account a possible temporal trajectory through data points (i.e. a temporal sequence of data points as it would result e.g. from electrophysiological measurements) as given in Figure 1h. Nevertheless, labeled clusters that form trajectories within a restricted area in space may be considered as reflecting attractor dynamics. In that view, centroid positions and diameters of clusters would reflect attractor basins (cf. Fig. 3b).
For statistical analysis of clusters we next permute the points' labels (random re-labeling) and re-compute the mean intra- and inter-cluster distances and the corresponding discrimination values (Fig. 1i). By that we estimate the cumulative distribution function of the discrimination values (Fig. 1j, k) from which we derive p-values to decide whether or not the analyzed clusters are significantly different from each other. Note that, in principle the described MCS method would equivalently well work with the 2-dimensional projected points (Fig. 1e) instead of the high-dimensional original data points, since firstly MDS is the only dimension reduction method that preserves all mutual distances and secondly our statistical method only relies on Euclidean distances (Fig. 1b,f). Nevertheless, we apply MCS to the original high-dimensional data and use MDS only as a tool to visualize the data points (Fig. 1e), temporal trajectories (Fig. 1h) and cluster proximities (Fig. 1g), reflecting relative positions and diameters of attractor basins (Fig. 1g).

\begin{figure}[htb!]
	\centering
	\includegraphics[width=1.0\linewidth]{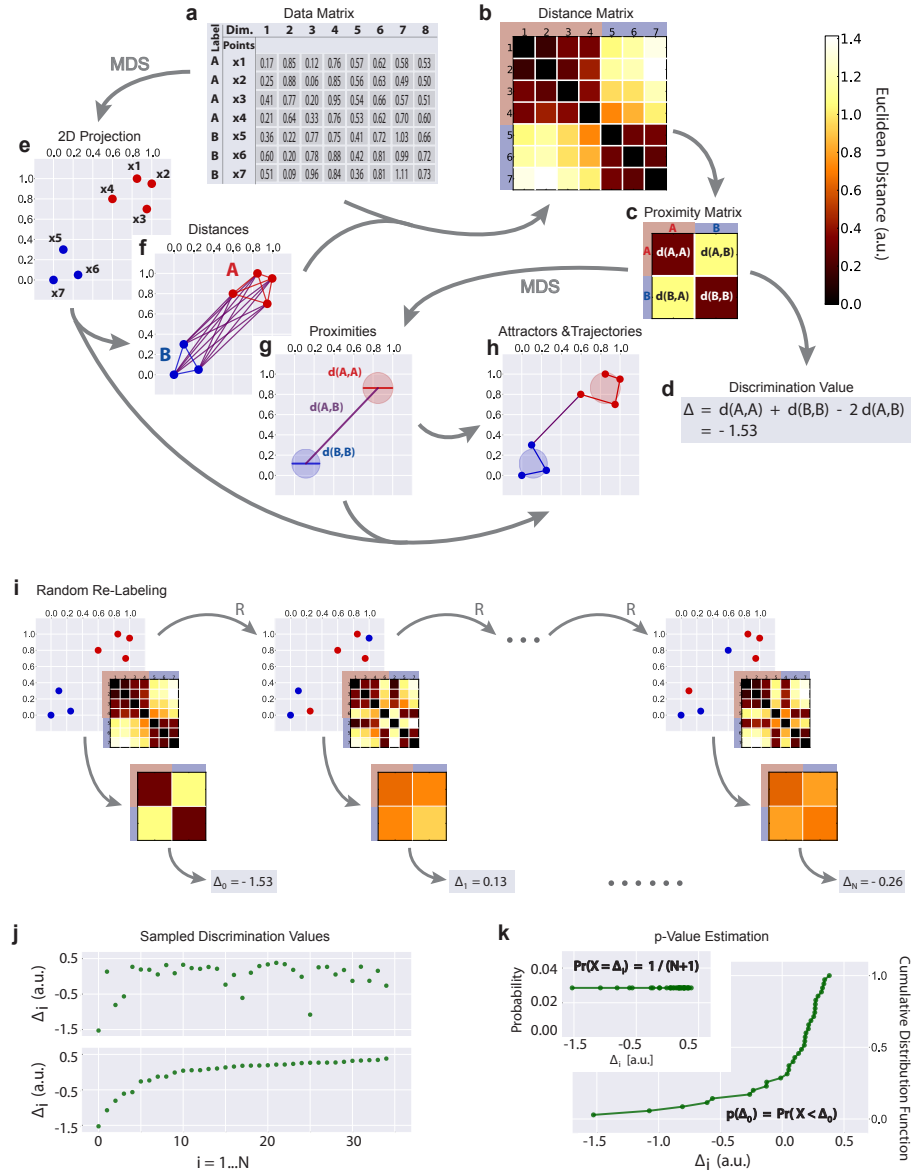}
	\caption{Comparing two clusters in n-dimensional space}
\end{figure}

\paragraph*{Comparing k clusters in n-dimensional space}
The method as described above for the statistical comparison of two clusters of data points in n-dimensional space can in principal be extended to k clusters in n-dimensional space. How this can be achieved is demonstrated in Figure 2 for three clusters A, B, and C, analogously to the procedure described in Figure 1: Again, 8-dimensional data were used (data matrix not shown) as visualized in Figure 2a. From these, again all pairwise Euclidean distances between points are calculated (Fig. 2b) and mean distances between intra- and all combinations of inter-cluster points are derived (Fig. 2c,d). From these the mean distances again discrimination values $\Delta$ can be computed for all pairwise cluster combinations as well as the combination of all three clusters (Fig. 2e). All further steps (random re-labelling of data points, cumulative distribution function, p-value estimation) are analogously based on these D values as described in Figure 1 for two clusters.

\begin{figure}[htb!]
		\centering
		\includegraphics[width=1.0\linewidth]{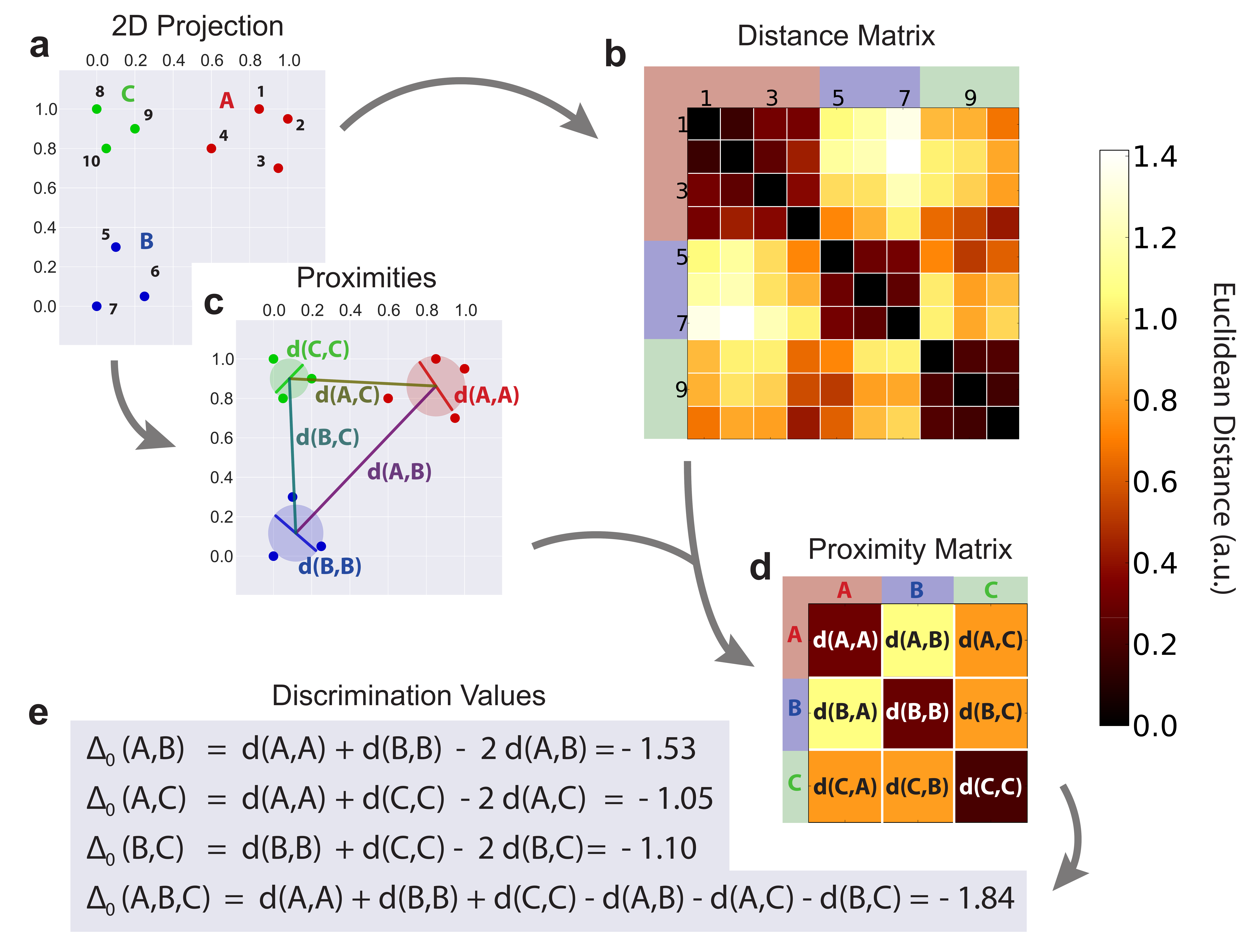}
		\caption{Comparing k clusters in n-dimensional space}
\end{figure}

\paragraph*{Analyzing steady-state LFP data from gerbil auditory cortex}
In order to apply MCS to electrophysiological multichannel LFP data, we consider the spatiotemporal neuronal activity pattern at a given time slot to be reflected by a data point in n-dimensional space, where each LFP recording channel provides amplitude values for one dimension. In other words, the LFP amplitude values within certain time windows for each recording channel can be viewed as n-dimensional vectors that represent the coordinates of a point in n-dimensional space, where n equals the number of recording channels. As a function of time, the succession of data points form a trajectory through n-dimensional space which reflects the dynamic changes of spatiotemporal activation patterns.
Figure 3 demonstrates the outcome of this approach for LFP data from gerbil auditory cortex. First, the mean LFP signal power within each recording channel was determined by moving a time window through the data (20s width, shifted in 5s steps) and calculating the root-mean-square (RMS) amplitude and z-transformation to remove activity common to the channels. By that, the amplitudes of the recorded potentials (power values) from each of the 16 recording channels within one time step were summarized into an 16-dimensional state vector (Fig. 3a, vertical grey bars) which thereby characterized the spatial distribution of neuronal activity. The temporal development of this spatial activity pattern corresponds to a trajectory in an abstract 16-dimensional state space. Recordings were made during 3 minute intervals of silence (red) and stimulation with pure tones of 1, 2 and 4 kHz (yellow, green, and blue, respectively); Note that during such sustained stimulation over several minutes, stimulus vs. no-stimulus conditions may not be distinguished based of neuronal discharge rate, that is, the sustained overall activity based on LFP amplitudes is not significantly different between conditions (cf. Fig. 3a). MCS therefore analyzes whether the LFP activity patterns across the 16 recording channels in our array were specific to certain stimulus conditions and may be distinguished from spontaneous activity during no-stimulus conditions, even seconds or minutes after stimulus onset.
To visualize the state vectors the corresponding data points in 16-dimensional space were projected onto a 2-dimensional target space by means of MDS which resulted in a trajectory on a 2-dimensional plane (Fig. 3b). Obviously, stimulus driven neuronal attractors are not only clearly separated from the one during silence (red) but also shift systematically with stimulation frequency. Note that the distance (reflecting dissimilarity of neuronal activity patterns) between the silence condition and each of the stimulus conditions is larger than between different stimulus conditions. Also note that when stimulus conditions were repeated (Fig. 3b, bottom panels) activity patterns jumped back to the field where they had been during the first respective presentation. This demonstrates that steady state LFP patterns are highly reliable and characteristic for a certain stimulus condition and by that show the attractor-like neuronal dynamics underlying these patterns. Independent from pre-stimulus activity patterns, reflected by different locations in state space and target space, the temporal trajectories always tend to converge to, and persist, within certain regions in state and target space, i.e. attractors that hence are characteristic to certain stimulus conditions (Fig. 3b, large colored dots). Using MCS it turned out that all attractors were highly significantly different from each other ($p<0.001$). These findings were consistent in all animals tested (n=15). Relative positions and diameters of attractor basins were computed from cluster proximities as explained above and visualized together with the projected trajectories. Figure 3c schows a summary of this analysis (overlay of all panels of Fig. 3b).
Note that the entire analysis of all measured data has been done with original 16-dimensional data and not with projected 2-dimensional data. Furthermore, we emphasize that, in contrast to Ohl et al., 2001, we do not search for peaks in any dissimilarity functions between different stimulus conditions (which are largely equivalent to onset responses) to pre-select only those states that are maximal different before analyzing or visualizing them but instead take into account all measured data (180 seconds duration per trial) and all trials. Thus the resulting differences of locations and diameters of clusters reflect differences in steady state responses and are not due to differences in the onset response.

\begin{figure}[htb!]
		\centering
		\includegraphics[width=1.0\linewidth]{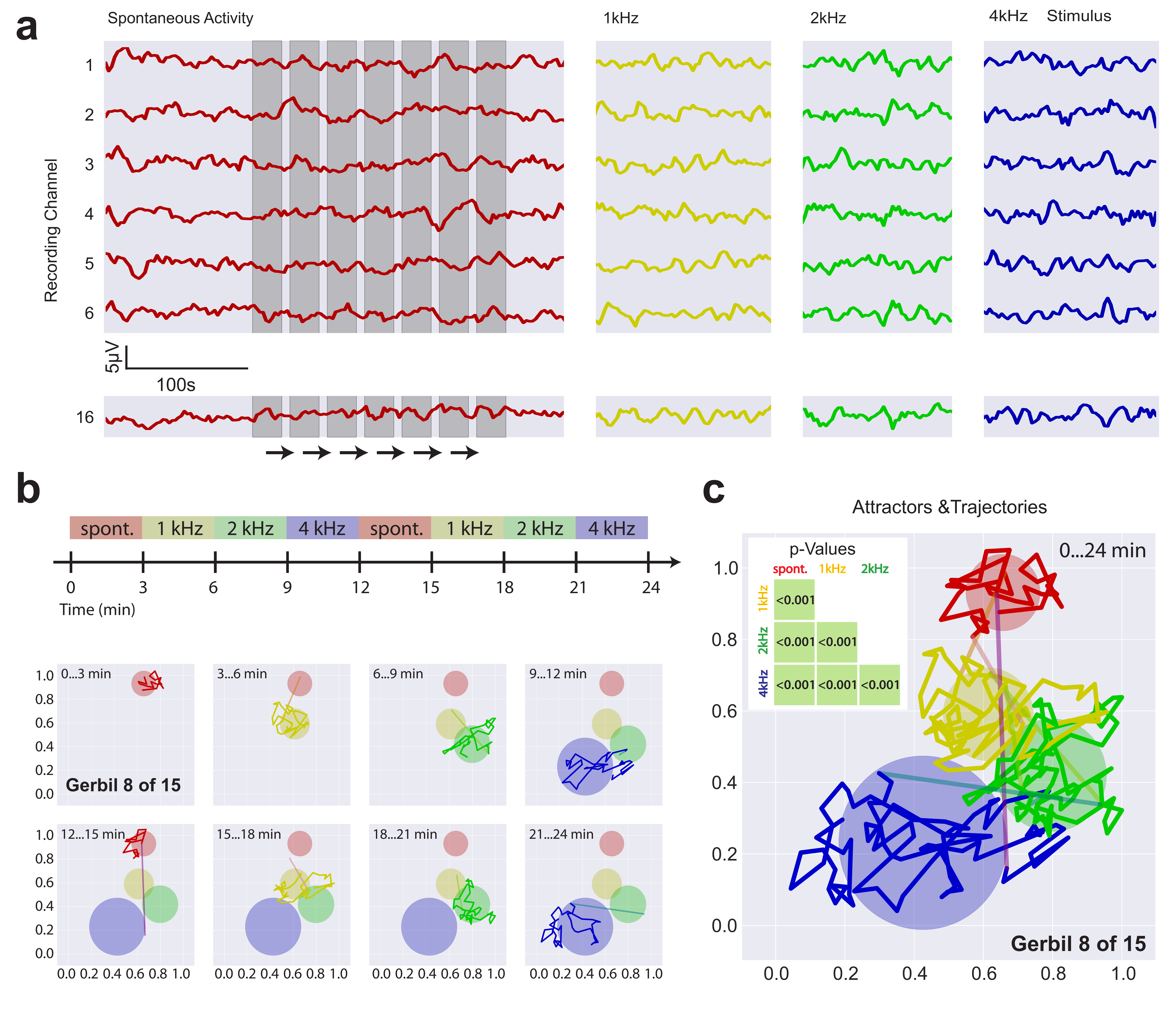}
		\caption{Analyzing steady-state LFP data from gerbil auditory cortex}
\end{figure}

\section*{Conclusion}
In this report, we present a new statistical method to analyze multichannel steady-state local field potentials (LFP) recorded within different sensory cortices of different rodent species. We demonstrate that using this approach stimulus-specific spatiotemporal activity patterns can be detected and be significantly distinguished from each other during stimulation with long-lasting stimuli. In future studies we plan to extend the method to human electroencephalogram (EEG) and magnetoencephalogram (MEG) data, demonstrating the universal applicability of our approach. Our method thereby may be used for the development of new read-out algorithms of brain activity and by that opens new perspectives for the development of brain-computer interfaces.

\section*{References}

Brody RM, Nicholas BD, Wolf MJ, Marcinkevich PB, Artz GJ (2013) Cortical deafness: a case report and review of the literature. Otol Neurotol 34:1226-1229.

Daelli V, Treves A (2010) Neural attractor dynamics in object recognition. Exp Brain Res 203:241-248.

Das A, Huxlin KR (2010) New approaches to visual rehabilitation for cortical blindness: outcomes and putative mechanisms. Neuroscientist 16:374-387.

Dehaene S, Sergent C, Changeux JP (2003) A neuronal network model linking subjective reports and objective physiological data during conscious perception. Proc Natl Acad Sci U S A 100:8520-8525.

Deliano M, Scheich H, Ohl FW (2009) Auditory cortical activity after intracortical microstimulation and its role for sensory processing and learning. J Neurosci 29:15898-15909.

Goldberg JM, Adrian HO, Smith FD (1964) Response of Neurons of the Superior Olivary Complex of the Cat to Acoustic Stimuli of Long Duration. J Neurophysiol 27:706-749.

Javel E (1996) Long-term adaptation in cat auditory-nerve fiber responses. J Acoust Soc Am 99:1040-1052.
Kumar A, Schrader S, Aertsen A, Rotter S (2008) The high-conductance state of cortical networks. Neural Comput 20:1-43.

Ohl FW, Scheich H, Freeman WJ (2001) Change in pattern of ongoing cortical activity with auditory category learning. Nature 412:733-736.

Ohl FW, Deliano M, Scheich H, Freeman WJ (2003a) Analysis of evoked and emergent patterns of stimulus-related auditory cortical activity. Rev Neurosci 14:35-42.

Ohl FW, Deliano M, Scheich H, Freeman WJ (2003b) Early and late patterns of stimulus-related activity in auditory cortex of trained animals. Biol Cybern 88:374-379.

Phillips DP, Orman SS, Musicant AD, Wilson GF (1985) Neurons in the cat's primary auditory cortex distinguished by their responses to tones and wide-spectrum noise. Hear Res 18:73-86.

Ringach DL (2009) Spontaneous and driven cortical activity: implications for computation. Curr Opin Neurobiol 19:439-444.

Salti M, Monto S, Charles L, King JR, Parkkonen L, Dehaene S (2015) Distinct cortical codes and temporal dynamics for conscious and unconscious percepts. Elife 4.

Schulze H, Ohl FW, Heil P, Scheich H (1997) Field-specific responses in the auditory cortex of the unanaesthetized Mongolian gerbil to tones and slow frequency modulations. J Comp Physiol A 181:573-589.

Thomas H, Tillein J, Heil P, Scheich H (1993) Functional organization of auditory cortex in the mongolian gerbil (Meriones unguiculatus). I. Electrophysiological mapping of frequency representation and distinction of fields. Eur J Neurosci 5:882-897.

Tomov P, Pena RF, Zaks MA, Roque AC (2014) Sustained oscillations, irregular firing, and chaotic dynamics in hierarchical modular networks with mixtures of electrophysiological cell types. Front Comput Neurosci 8:103.

%\bibliographystyle{plain}

%\bibliographystyle{unsrt}
%\bibliography{refs}

\end{document}